\documentclass[aps,prb,twocolumn,showpacs,amsmath,amssym,
superscriptaddress]{revtex4-1}
\usepackage{amssymb}
\usepackage{graphicx}
\usepackage{dcolumn}
\usepackage{color}
\usepackage{verbatim}
\begin{document}

\title{First-principles study of multiferroic RbFe(MoO$_4$)$_2$}

\author{Kun Cao}

\email{kun.cao@materials.ox.ac.uk}
\affiliation{Key Laboratory of Quantum Information, University of
Science and Technology of China, Hefei, 230026, People's Republic of
China}
\affiliation{Department of Materials, University of Oxford, Parks Road, Oxford OX1 3PH, United Kingdom}

\author{R. D. Johnson}
\affiliation{Clarendon Laboratory, Department of Physics, University of Oxford, Parks Road, Oxford OX1 3PU, United Kingdom}
\affiliation{ISIS Facility, STFC-Rutherford Appleton Laboratory, Didcot, OX11 OQX, United Kingdom}

\author{Feliciano Giustino}
\affiliation{Department of Materials, University of Oxford, Parks Road, Oxford OX1 3PH, United Kingdom}

\author{Paolo G. Radaelli}
\affiliation{Clarendon Laboratory, Department of Physics, University of Oxford, Parks Road, Oxford OX1 3PU, United Kingdom}

\author{G-C Guo}
\affiliation{Key Laboratory of Quantum Information, University of
Science and Technology of China, Hefei, 230026, People's Republic of
China}

\author{Lixin He}
\email{helx@ustc.edu.cn}
\affiliation{Key Laboratory of Quantum Information, University of
Science and Technology of China, Hefei, 230026, People's Republic of
China}

\date{\today}

\begin{abstract}
We have investigated the magnetic structure and ferroelectricity in RbFe(MoO$_4$)$_2$ via first-principles calculations. Phenomenological analyses have shown that ferroelectricity may arise due to both the triangular chirality of the magnetic structure, and through coupling between the magnetic helicity and the ferroaxial structural distortion. Indeed, it was recently proposed that the structural distortion plays a key role in stabilising the chiral magnetic structure itself. We have determined the relative contribution of the two mechanisms via \emph{ab-initio} calculations. Whilst the structural axiality does induce the magnetic helix by modulating the symmetric exchange interactions, the electric polarization is largely due to the in-plane
spin triangular chirality, with both electronic and ionic contributions being of relativistic origin. At the microscopic level, we interpret the polarization as a secondary steric consequence of the inverse Dzyaloshinskii-Moriya mechanism and accordingly explain why the ferroaxial component of the electric polarization must be small.

\end{abstract}
\maketitle


\section{Introduction}

Magnetic ferroelectrics, in which ferroelectricity is induced by magnetic ordering,
have attracted great attention recently for their novel physics and potential
device applications.\cite{cheong07,fiebig05,hur04b,kimura03,goto04,hur04} 
There are several mechanisms that can induce electric polarization upon
magnetic ordering in these materials,
such as exchange striction,\cite{wang07,wang08} the KNB mechanism\cite{katsura05}
and the inverse DM mechanism. \cite{sergienko06} These theories
explain very well the experimental observations in, for example, the canonical multiferroics TbMn$_2$O$_5$, TbMnO$_3$
and many other compounds. \cite{chapon04,radaelli09,hur04}
Recently, a new class of magnetoferroelectric materials has been
discovered in which the electric polarization is 
\emph{perpendicular} to the spin rotation plane, \cite{johnson11,johnson12,hearmon12} and cannot be explained by the aforementioned mechanisms.
In particular, the KNB theory predicts that 
the polarization should lie in the plane of rotation of the spins.
However, in some cases this perpendicular polarizaition {\bf P} can be induced by the helicity 
$\sigma$ of the
magnetic structure in a class of materials labelled 
ferroaxials. \cite{johnson11} In these materials,  
there exists a macroscopic structural axial vector ${\bf A}$, representing a global rotation in the crystal structure, 
which is only allowed in seven point group symmetries;
$\overline{1}, 2/m, \overline{3}, \overline{4}, \overline{6}, 4/m, 6/m$.

\begin{figure}
\centering
\includegraphics[width=3.5in]{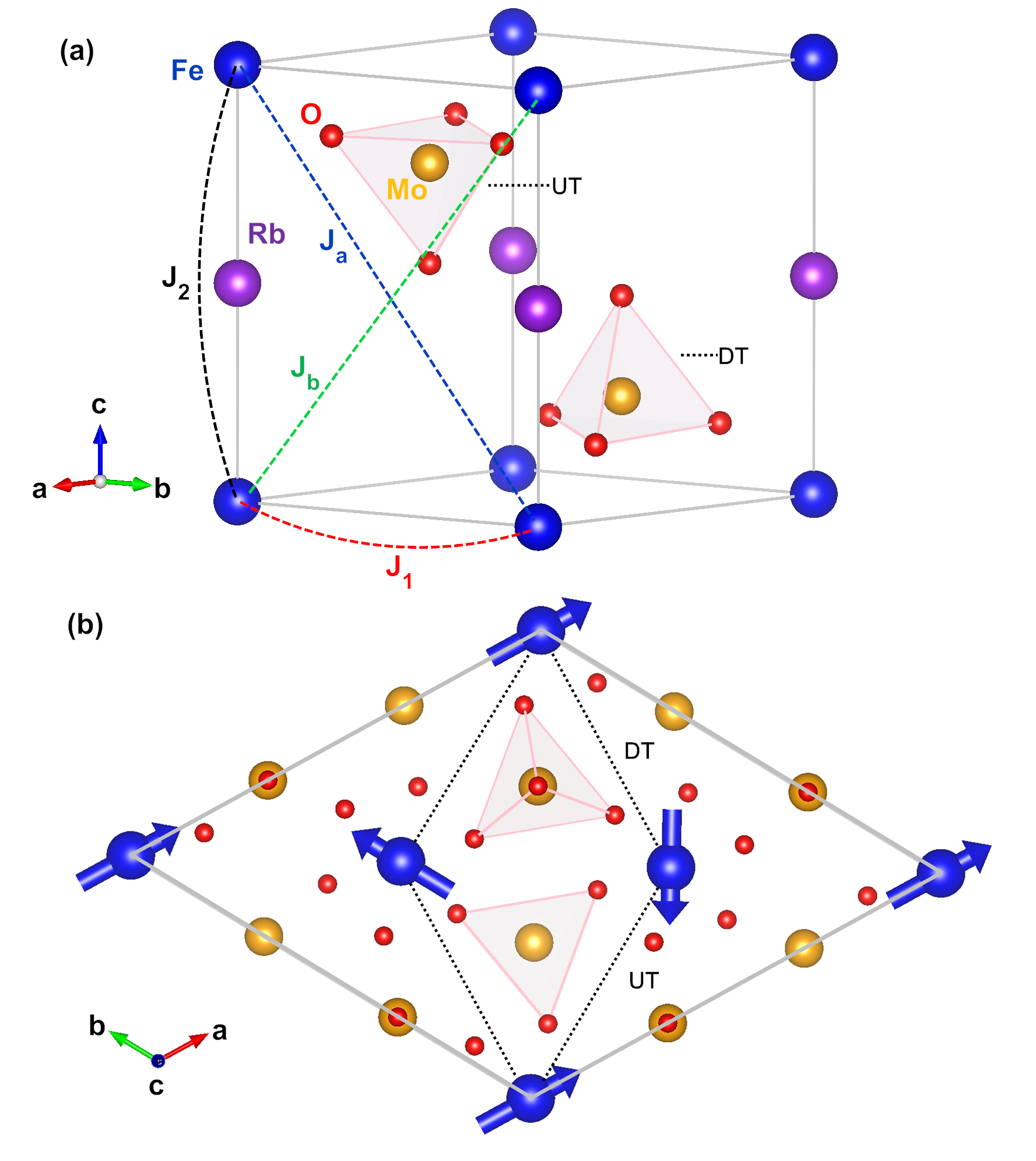}
\caption {A sketch of the lattice and spin strucures of RFMO.
(a) The unit cell of RMFO with exchange interaction paths marked. 
(b) The in-plane $120^\circ$ spin structure, shown in a $\sqrt{3} \times \sqrt{3}$
  supercell. Blue arrows denote 
the spin directions of Fe ions, whereas the black dashed lines mark one unit cell.}
\label{fig:structure}
\end{figure}

RbFe(MoO$_4$)$_2$, referred to as RFMO hereafter, is an obvious candidate to investigate this phenomenon, since it is ferroaxial at low temperatures (point group $ \overline{3}$) and has a particularly simple atomic structure and magnetic exchange path ways. \cite{hearmon12}
The RFMO crystal structure consists of an alternate stacking of a magnetic Fe$^{3+}$ layer, 
two MoO$_{4}$ layers (Mo$^{6+}$ in the centre of O$_4$ tetrahedra) and 
one Rb$^{1+}$ layer, as shown in Fig.~\ref{fig:structure}(a). 
Below T$_s$=190K, RFMO undergoes a structural transition with the MoO$_{4}$ tetrahedra 
rotating collectively around the $c$ axis. The lattice symmetry is therefore lowered from
$P\overline{3}m1$ above T$_s$, to ferroaxial $P\overline{3}$ below. 
When the temperature is reduced below $T_N=4$K, iron spins become magnetically
ordered as shown in
Fig.\ref{fig:structure}(b).  Neutron scattering experiments \cite{kenzelmann07} show that all spins lie in the $ab$ plane and form, in each layer, the $120^{\circ}$  structure typical of triangular lattices. Spins in subsequent layers along the $c$ axis are rotated by $\sim$ 158$^\circ$, defining an overall incommensurate helical envelope. The magnetic propagation vector corresponding to these two modulations is {\bf q}=(1/3, 1/3, 0.44). 
At the same time, a small spontaneous polarization arises at $T_N$ along 
the $c$ axis with  $P_c\approx 6 \mu$ C/m$^{2}$. \cite{kenzelmann07}  Kenzelmann {\it et al} [\citenum{kenzelmann07}] also proposed, based upon a phenomenological model, that free energy of RFMO could be described in terms of $\sigma_1^2$ and $\sigma_2^2$, where $\sigma_1$ and $\sigma_2$ represent the two structures with opposite in plane magnetic triangular chirality. According to this model, the structural distortion below T$_s$ determines the sign of $q_z$,  
while leaving the degeneracy between structures with opposite chirality
$\sigma_1$ and $\sigma_2$ unbroken. Following a similar path, they argued that the magnetoelectric polarization $P_c \propto \sigma_1^2-\sigma_2^2$.  Kaplan {\it et al} also proposed an interpretation for magnetoelectric coupling in RFMO based on general symmetry arguments.\cite{kaplan11}  These papers clearly showed that the triangular magnetic ordering is in itself sufficient to lower the symmetry to a polar point group, even in the absence of either ferroaxial distortion or magnetic helicity.

In a previous experimental paper,\cite{hearmon12} some authors of this paper proposed that the helical magnetic structure structure in RFMO, the origin of which remained unclear, is in fact induced by the structural axiality through symmetric exchange.  We also show that the axial distortion can give rise to a second component of the electrical polarization, so that the total polarization can be written phenomenologically as $P_c=c_1\sigma_t+c_2  A \sigma_h$, 
where $c_1$ and $c_2$ are constants, $\sigma_t$ represents the in plane triangular chirality, $\sigma_h$ 
is the magnetic helicity along the $c$ direction, and $A$ is the component of the axial rotation parallel to the electric polarization (in RFMO, the ferroaxial vector is ${\bf A}= (0, 0, A)$, hence
$A$ is the magnitude of the ferroaxial distortion whose sign is determined by the structural rotation direction). Both terms are of antisymmetric origin and have the same symmetry properties in the space group $P\bar{3}$, but imply different microscopic mechanisms.  The first term describes the component of the polarisation induced by the triangular chirality of spins in the plane, which would be present even for $A=0$, as described in reference\citenum{kenzelmann07}.  If one was able to tune  $A$, i.e., the rotation angle of the MoO$_4$ tetrahedra, for example by applying an external pressure, one would expect a \emph{linear} behaviour of the polarization with non-zero intercept as a function of $A$. Indeed, there is no \emph{prima facie} reason why the ferroaxial term should be small.  The spin rotation away from collinearity ($\sim$22$^\circ$) is of the same order of that of TbMnO$_3$, the prototypical magnetic multiferroic. One might expect, therefore, that coupling to $\sigma_h$ in RFMO may be significant. Furthermore, in RFMO the spin rotation arises due to the symmetric exchange interaction, and is not a small effect of relativistic origin. In addition, the axial rotation gives rise to large atomic displacements of $\sim$ 0.1\AA, much larger than any ferroelectric displacement. Hence, a substantial ferroaxial component cannot be ruled out on either symmetry or phenomenological grounds.

In this paper, we perform first-principles calculations to study the
electronic and magnetic structures of RFMO, and to clarify the origin of the
electric polarization in this material. We demonstrate that the ferroaxial structure
in RFMO does indeed induce the incommensurate helical spin structure
along the $c$ direction via the symmetric exchange interactions. 
The electric polarization in RFMO is interpreted as a secondary effect of a structural distortion induced through the inverse DM mechanism. We find the electric polarization resulting from the ferroaxial mechanism to 
be negligible (i.e., $c_2 \approx$0) compared to the contribution from the triangular chirality. We further explain why the ferroaxial contribution
in this material should be small.

\section{Methods}

Our first-principles calculations are based on density-functional theory
implemented in the Vienna ab initio simulations package
(VASP).\cite{kresse93,kresse96} 
We use the spin-polarized generalized gradient approximation 
with on-site Coulomb interactions U included for Fe 3d
orbitals (GGA+U). \cite{liechtenstein95}
Although several $U$ values were tried, here we present the results for $U$=4 eV and 6 eV and $J$=0.9 eV, which are typical for
iron.
Calculations were based on the experimental structure, but a $\sqrt{3} \times \sqrt{3}\times 3$  supercell was used, corresponding to a commensurate wave  vector ${\bf q}_m=(1/3, 1/3, 1/3)$, which is a reasonable approximation to
the experimental wave vector ${\bf q}_m=(1/3, 1/3, 0.44)$. The spin-orbit coupling (SOC) was taken into account in the calculation unless otherwise stated.  We used the projector augmented-wave (PAW)\cite{blochl94} method with a 400 eV
plane-wave cutoff.   A $2\times 2 \times 1$ k-points mesh gave good convergence.  We relaxed the crystal structure until the changes of total energy in the self-consistent calculations were less than 10$^{-7}$ eV and the remaining forces are less than 1 meV/\AA.

\section{Results and discussion}

\subsection{Magnetic structures}

Since the magnetic structure of this material forms an incommensurate spiral
along the $c$ direction, it is difficult to study directly the exact experimental magnetic structure from first-principles
calculations. Therefore, we fitted our first-principles calculated energies to a Heisenberg model Hamiltonian, including both 
nearest neighbour (NN) and next nearest neighbour (NNN) exchange interactions.
The four exchange interactions are shown in Fig.~\ref{fig:structure},
where $J_1$ is the in-plane NN exchange interaction, $J_2$ is exchange interaction along
the $c$ direction and $J_a$, $J_b$ are NNN exchanges along diagonal paths.
The exchange energy per Fe ion of the experimentally observed
magnetically ordered state can be written as,
\begin{eqnarray}
\label{eq:energy}
E &=& -\frac{3}{2}J_1+J_2\cos(2\pi q_z)-\frac{3}{2}(J_a+J_b)\cos(2\pi q_z) \\ \nonumber 
&& +\frac{3\sqrt{3}}{2}(J_b-J_a)\sigma_h\sigma_t \sin(2\pi q_z)\, ,
\end{eqnarray}
where $\sigma_t$ and $\sigma_h$ represent in-plane and out of plane
chiralities respectively.
For example, ${\bf q}_m =(1/3,1/3, \pm 1/3)$ has
  $(\sigma_t, \sigma_h)=(+1, \pm 1)$, whereas ${\bf q}_m=(-1/3,-1/3,\pm 1/3)$ has
$(\sigma_t, \sigma_h)=(-1, \pm 1)$ etc.
By minimizing the magnetic energy with respect to  $q_z$, we obtain:
\begin{equation}
\tan(2\pi q_z)=\frac{3\sqrt{3}(J_a-J_b)}{3(J_a+J_b)-2J_2} \, .
\label{eq:qz}
\end{equation}
It can be seen from Eq.~\ref{eq:qz} that if $J_a \neq J_b$, $q_z \neq 0$, and a
spin helix can be formed along the $c$ direction to minimize the exchange energy. 

\begin{figure}
\centering
\includegraphics[width=3.3in]{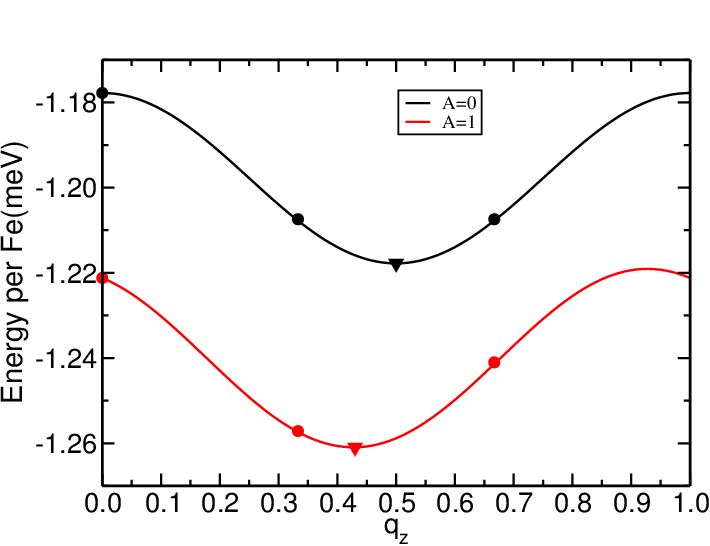}
\caption {The magnetic energies as functions of $q_z$ 
for structures with A=1 and A=0.
The in plane spin configurations are arranged according to
Fig. \ref{fig:structure}(b).  
Dots represented the energies from first-principles calculations 
with $U$=4eV, whereas
the solid lines are fitted using Eq.\ref{eq:energy}. 
The triangles mark the energy minima.}
\label{fig:energy}
\end{figure}

To determine the exchange interactions $J_1$, $J_2$, $J_a$ and $J_b$, we fit the
total energy calculated from first principles to a Heisenberg model
 for both cases of $A$=1 (ferroaxial, below $T_s$) and $A$=0 (non-ferroaxial above  $T_s$).
The total energy was calculated using five spin configurations in the $\sqrt{3} \times \sqrt{3} \times 3$
super-cell, including ${\bf q}_m=(1/3, 1/3, \pm 1/3)$, and ${\bf q}_m=(1/3, 1/3, 0)$, and two collinear spin configurations,
($\sigma_t=+1,\upuparrows \downarrow$), ($\upuparrows \downarrow,\sigma_h=0$),
where $\sigma_h=0$ refers to $q_z$=0, and
$\upuparrows \downarrow$ represents up-up-down configuration within the $ab$ plane or along the $c$ direction. The calculated exchange interactions are reported in Table \ref{tab:exchange}.
Having obtained the exchange interactions, we calculated $q_z$ according to Eq.\ref{eq:qz}.  
For the non-ferroaxial structure with $A$=0, $J_a=J_b$ is enforced by symmetry, while for the ferroaxial structure, $ A \neq 0$, $J_a$ and $J_b$ are no longer equivalent. Figure~\ref{fig:energy} depicts total energies calculated from first principles
(in dots) for ${\bf q}=(1/3,1/3,q_z)$, with $q_z$=0, 1/3, 2/3 for both $A$=0 and $A$=1 using $U$=4 eV, and the energy curves (solid lines) fitted 
using the Heisenberg model in Eq.\ref{eq:energy}. For the non-ferroaxial structure ($A$=0) the energy minimum lies at $q_z=0.5$, so there is no incommensurate helix, whereas the energy minimum lies at $q_z$=0.43 for the ferroaxial structure with $A$=1, which is in excellent agreement with the experimental value $q_z=0.44$. 
It is worth noting that although the specific values of
    $J_a$ and $J_b$ depend on the Coulomb $U$ used in the calculations, as shown in Table \ref{tab:exchange}, 
   the ratio $(J_a-J_b)/J_2$ for $A=1$ is robust.  The reason is as
   follows. When $U$ increases, the electrons become more localized, and the exchange energies 
  $J_2$, $J_a$, $J_b$ decrease simultaneously in an equal manner. 
  Therefore the ratio $(J_a-J_b)/J_2$ shows no significant variation with $U$, giving a robust determination of the wave vector $q_z$.
Furthermore, by symmetry the sign of
  $(J_a-J_b)$ is determined by the sign of A. 
The last term of Eq.\ref{eq:energy} can therefore be re-written as $\propto A \sigma_h\sigma_t$ which governs the ground state magnetic domain structure, \emph{i.e.} the helicity $\sigma_h$ of the stable structure in each ferroaxial domain switches sign if the sign of the triangular chirality $\sigma_t$ is reversed.
This represents the first important result of our analysis:  the ferroaxial rotation of the MoO$_4$
tetrahedra induces the incommensurate helical spin structure along the $c$ axis, and the sign of the helicity is determined by the product $A\sigma_t$.

We also performed Monte carlo (MC) simulations to determine the magnetic
transition temperature $T_N$, and the results are listed in Table
\ref{tab:exchange}.
The simulations indicate that $T_N$ is mostly determined by the $J_1$ and
$J_2$ interactions, whereas $J_a$ and $J_b$ are too small to affect the
transition temperature significantly.
The exchange interactions fitted with $U$=4 eV give $T_N$=3.3 K, 
which is in good agreement with the experimental value $T_N$=4 K.
It is worth noting that the calculated exchanges interactions with $U$=4 eV are also in good agreement with 
recent results fitted from experimental data.\cite{white13}    

\begin{table}
\begin{center}
\tabcolsep 1mm
\caption{The calculated exchange interactions, $q_z$ and $T_N$. The exchange interactions are in the unit of meV. }
\begin{tabular}{ccccccccc}
\hline \hline
U(eV) &  A &   $J_1$   & $J_2 $   & $J_a$ & $J_b$  &$\frac{J_a-J_b}{J_2}$& $q_z$ & $T_N$(K) \\
\hline
0 & 1   & 3.66          & 0.042              & -0.0050   &  0.0040  & 0.21   & 0.42 & 16 \\
4 & 0   & 0.80         & 0.019               & -0.0004  &  -0.0004  &  0  & 0.50  \\
4 & 1   & 0.76          &0.018               & -0.0020   &  0.0014  &  0.19  & 0.43 & 3.3 \\
6 & 1   & 0.27          & 0.010               & -0.0010   & 0.0010  &   0.20 & 0.42 & 1.2\\
\hline
\end{tabular}
\label{tab:exchange}
\end{center}
\end{table}

\subsection{Electric polarization}

In magnetic ferroelectrics, the electric polarization can usually be decomposed in a pure electronic contribution and an ionic
contribution. \cite{wang07,lottermoser09}
We first study the pure electronic contribution to the polarization, by performing 
the calculations while constraining the ions to the high symmetry positions of space group P$\overline{3}$ . The iron spins were also constrained to lie in the $ab$ plane with ${\bf q}_m=(1/3, 1/3, 1/3)$.
In magnetic multiferroics, the electric polarization calculated using the
GGA+U scheme is usually very sensitive to the on-site Coulomb U parameters. \cite{wang07,giovannetti08}  We therefore performed the calculations for several different U parameters. 
In all cases the calculated polarisation was along the $c$ axis, and the calculated values were $P_c= 11 \mu C/m^{2}$ and $P_c=7 \mu C/m^{2}$ for U= 4 eV and for U=6 eV, respectively, which are of the same order of the experimental value. 
We also carry out calculations in the absence of SOC, and the calculated $P$ vanishes, indicating that the electric polarization is entirely due to the antisymmetric exchange.
To extract the respective contribution from the two different terms $c_1\sigma_t$ and $c_2A\sigma_h$,
we further calculate $P$ for the same $\sigma_t$ but with $q_z$=-1/3, $q_z$= 0 and  $q_z$= 1/2. 
Interestingly, the calculated $P$ remains almost unchanged with different
$q_z$. This suggests that the spin helix along the $c$ axis does not
directly contribute to the polarization. 
Furthermore, when the in-plane chirality $\sigma_t$ is reversed, the $P$ 
also changes sign, further confirming that 
the electric polarization only depends on $\sigma_t$. 
We also repeated the calculations with high symmetry geometry ($A$=0) and
find that the electric polarization does not change. This insensitivity of polarization to out-of-plane helicity is also proposed by  Kenzelmann {\it et al}. \cite{kenzelmann07}
Therefore, according to our calculation, the polarization in RFMO is entirely
determined by the triangular chirality of the ion spins in the plane, i.e., $P_c=c_1\sigma_t$. 
This is consistent with our previous spherical neutron
polarimetry experiments \cite{hearmon12}, where the chiral domain population could be reversed by an applied electric
field, so that flipping $P$ always resulted in a simultaneous flip of $\sigma_t$.  
Even though the magnetoelectric coupling $c_2A\sigma_h$ is allowed by symmetry, our results indicate that this term is very small in RFMO. 
In spite of this, $\sigma_h$ is also expected to flip by flipping $P$ in each ferroaxial domain, since the symmetric-exchange coupling term $A \sigma_h \sigma_t$ is large, as shown in the previous section.  This is also consistent with the experiment.

\begin{figure}
\centering
\includegraphics[width=3.5in]{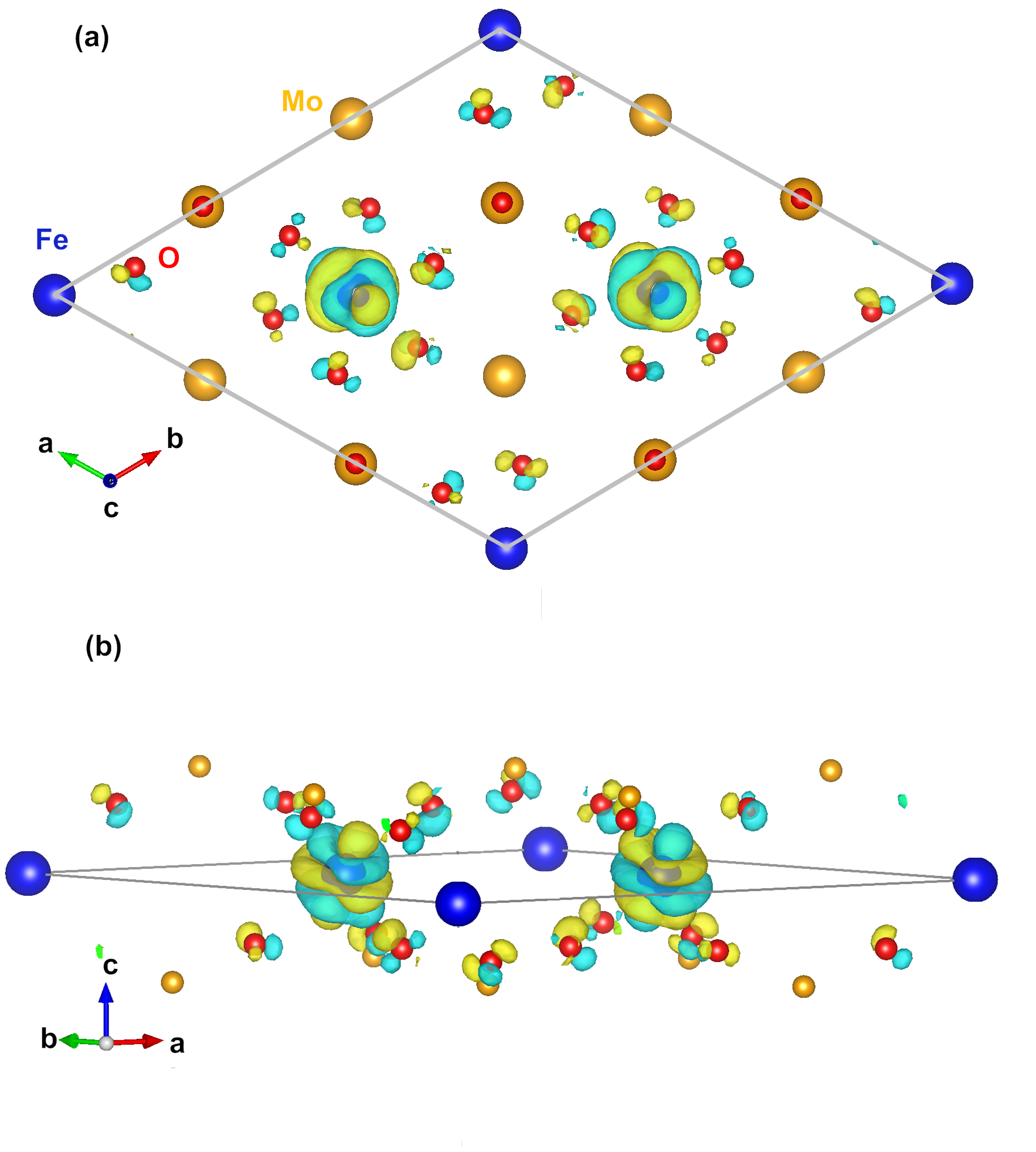}
\caption {Plot of the treated charge density $\delta n(x,y,z)$. Only the charge density around Fe(dark blue) and O(red) is plotted. 
The blue isosurfaces denote n(x,y,z) with + sign, while the yellow isosurfaces denote - sign.}
\label{fig:charge}
\end{figure}

To understand the origin of the purely electronic component of the polarization, 
we calculated site-projected charges (SPCs) and then produced a difference map of the SPCs between the two ions that are 
related by the inversion symmetry operation.
These values, defined as $\delta n(x,y,z)=n(x,y,z)-n(-x,-y,-z)$, where $n(x,y,z)$ is the
charge density at $(x,y,z)$, can be reasonably interpreted as charge transfer (CT),
which break the inversion symmetry of charge density. 
Without SOC, the integrated CTs of each single pair of symmetry-related sites are of the order of $10^{-6} e$ 
and oscillate with alternating signs, so the total CT is very small and of the order of  $10^{-7} e$.
When including SOC, the CTs are much larger, of the order
of $10^{-5} e$ for a single pair of atoms, while the overall CT is approximately $1.5 \times 10^{-4} e$. 
The CT maps calculated in the present of SOC are plotted in Fig.\ref{fig:charge} in order to visualise the charge transfer distribution in more detail.  Significant charge transfer occurs for Fe and O atoms, which was found to be predominantly along the $c$ axis, and results in the development of the electric polarization. 

\begin{figure}
\centering
\includegraphics[width=3.3in]{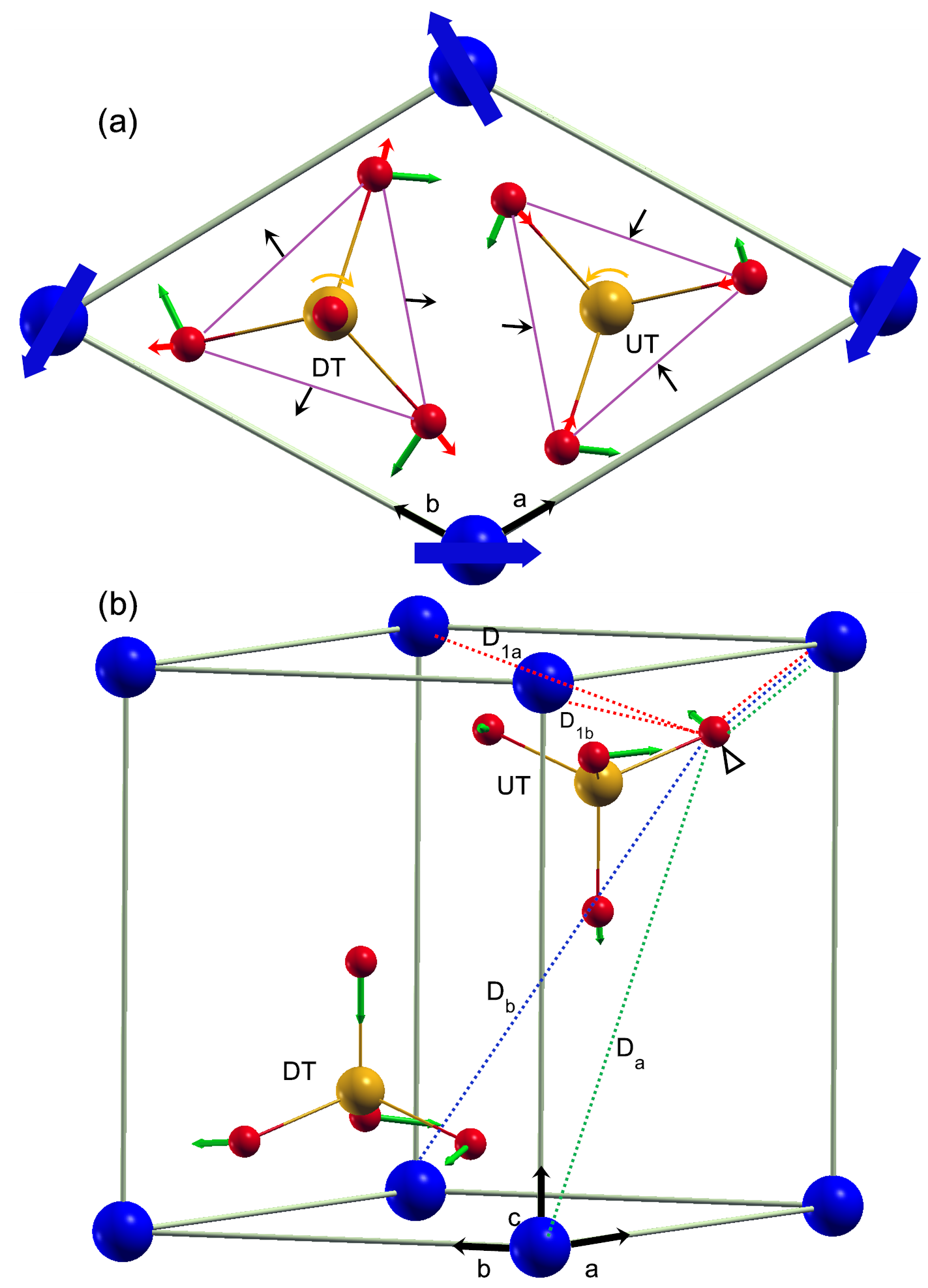}
\caption {Displacement pattern for the two tetrahedra in the unit cell, calculated with
  U=0. Green arrows represent calculated displacements. (a) Red arrows denotes the
  components of the calculated displacement which provide the stretch and expansion 
  effect. Blue arrows denote spin directions. Black arrows denote local magnetic polarity vectors (see text). Yellow arrows mark the rotation direction of tetrahedra. 
(b) All the relevant DM interactions  pertaining to one O atom located in the basal plane of the top MoO$_4$ tetrahedron are marked.}
\label{fig:disp}
\end{figure}

\begin{table}
\begin{center}
\tabcolsep 2.5mm
\caption{The O atomic positions in fractional coordinates and calculated displacement with U=0 in cartesian coordinates of $10^{-5}$\AA.}
\begin{tabular}{cccrrr}
\hline \hline
          & positions &          &          &    disp  &  \\
\hline
   0.33333 &  0.66667 &  0.53683 &    -0.9  &    -0.4  &   -11.0 \\
   0.66667 &  0.33333 &  0.46286 &     0.1  &     0.0  &   -21.9 \\
   0.78200 &  0.67900 &  0.15795 &   -10.9  &    23.9  &    -3.6 \\ 
   0.32100 &  0.10300 &  0.15795 &   -15.4  &   -21.3  &    -3.6  \\
   0.89700 &  0.21800 &  0.15795 &    26.2  &    -2.5  &    -3.6  \\
   0.21800 &  0.32100 &  0.84174 &   -12.2  &    17.0  &     3.3  \\
   0.67900 &  0.89700 &  0.84174 &   -14.0  &   -18.3  &     2.2  \\
   0.10300 &  0.78200 &  0.84174 &    15.9  &    -4.0  &     5.1  \\
\hline
\end{tabular}
\label{tab:disp}
\end{center}
\end{table}

We now examine the impact of ion relaxation on the electric polarization, by relaxing the structural geometry in the spin configuration ($\sigma_t = 1, \sigma_h=0$) using a $\sqrt{3} \times \sqrt{3}\times 1$ supercell. 
Since the magnetic configuration preserves the 
three fold rotation symmetry, the final relaxed structure has space group
$P3$. The relaxed structure calculated with U=6 eV has a total polarization $P_c=14
\mu C/m^{2}$, indicating that both electronic and ionic effects contributed to $P$.
To determine the ferroelectric distortion due to spin-lattice coupling 
in this material, we first construct a high symmetry structure with space 
group $P\bar{3}$ by averaging the relaxed structure between the two directions of $P$. \cite{wang07,cao09}
The atomic displacements are then obtained by calculating the difference between the relaxed and averaged structures.  We have also repeated the same calculations in the absence of axial rotation (space groups P$\overline{3}$m1 $\rightarrow$ P$3$m1), and with
different U parameters. All the calculated displacement patterns are very similar, with the most significant
associated with oxygen atoms. The amplitude of atomic displacements calculated with U=6 eV is smaller than 10$^{-4}$ \AA. Although this amplitude of magnetoelectric distortion might be too weak to be reliably compared with experiments, the key features of the displacement pattern are robust during our calculations with various parameters. Therefore, to better illustrate the displacement pattern without losing generality, we show the much stronger pattern calculated with $U=0$ in Table.\ref{tab:disp} and Fig.~\ref{fig:disp}. 
Looking at the distortion in greater detail (Fig.~\ref{fig:disp}), one can decompose the displacement pattern in three components, all acting oppositely on the two tetrahedra within a unit cell (denoted in the figure as up tetrahedron, UT, and down tetrahedron, DT):  a \emph{rotation} of the tetrahedon clockwise/counterclockwise, the aforementioned displacements of the apical/basal oxygens and an expansion/shrinkage of the basal oxygen triangles.  Microscopically, the relevant mechanism is the antisymmetric
Dzyaloshinskii-Moriya (DM) exchange, where energy can be gained by distorting
the crystal structure.   When two magnetic ions carrying non-collinear spins are connected 
by common ligand atoms, the DM exchange energy between these two spins can be written as 
$E_{DM}={\bf D}\cdot ({\bf S}_1 \times {\bf S}_2)=\gamma ({\bf e}_{12}\times {\bf u}) \cdot ({\bf S}_1 \times {\bf S}_2)=-\gamma {\bf u} \cdot \left[{\bf e}_{12} \times ({\bf S}_1 \times {\bf S}_2)\right]$, 
where ${\bf S}_1$ and ${\bf S}_2$ are the spins on the two magnetic ions,
${\bf D}$ is the DM vector, ${\bf e}_{12}$ is the position vector connecting them,
${\bf u}$ is a position vector for the ligand, and $\gamma$ is a coupling
constant. The local magnetic polarity vectors $\left[{\bf e}_{12} \times ({\bf S}_1 \times {\bf S}_2)\right]$ are indicated with black arrows in Fig.~\ref{fig:disp}.   Energy can be gained by displacing the ligands, so that
\begin{equation}
\label{eq:  DMenergy}
 \Delta {\bf u} \propto \partial E_{DM}/\partial {\bf u}  \approx \gamma {\bf e}_{12}\times ({\bf S}_1 \times {\bf S}_2) \nonumber
  \end{equation}
where, for simplicity we ignore the term in $\partial \gamma/\partial {\bf u}$, which is expected to be small.  Each oxygen atom belongs to two  Fe-O-Fe clusters, and the total displacement can be considered as the vector sum of the two displacement vectors for each cluster.

The first set of displacements (tetrahedral rotations) disappears completely when the axial rotation is removed, and can be easily explained with the fact that, in P3m1, the DM forces on a given oxygen atom originating from the two clusters are equivalent by symmetry, whereas this is no longer the case in P3.  This imbalance necessarily generates a rotation of the tetrahedra, as shown in Fig.  \ref{fig:disp}.  The expansion/shrinkage of the basal oxygen triangles is unaffected by the presence of the axial distortion, and is the primary consequence of the inverse DM effect:  by expanding/contracting the basal triangles, the system decreases/increases the magnitude of the DM vector, thereby reducing its DM energy. Notably, these two sets of displacements can not directly contribute to the macroscopic polarization since their overall effect are cancelled by the three-fold rotation symmetry.  Finally, the displacement of the apical/basal oxygens along the $c$ axis (also unaffected by the axial distortion), which directly contributes to polarization, can be understood as a secondary steric effect that minimises the change in the volume of the tetrahedra caused by the expansion/shrinkage of the basal oxygen triangles. 

The above analysis, only including in-plane DM interactions, can be extended to the out of plane DM interactions
corresponding to the exchange paths of $J_a$ and $J_b$. All the exchange paths
related to one O atom located in the basal plane are illustrated in
Fig.\ref{fig:disp}(b), where ${\bf D}_{1a}$ and ${\bf D}_{1b}$ correspond to
in-plane DM interactions, ${\bf D}_{a}$ and ${\bf D}_{b}$ corresponds to
out of plane DM interactions. The coupling constants $\gamma$ of these DM interactions are denoted using the same subscripts as ${\bf D}$.
Taking all these DM interactions into account and assuming that the polarization is
proportional to the stretch forces exerted to O atoms, we can write an
expression for polarization,
\begin{eqnarray}
\label{eq:polar}
P_c & \propto & \gamma_{1a} \lambda_a \sin(\frac{\pi}{3} \sigma_t)+\gamma_{1b} \lambda_b \sin(\frac{\pi}{3} \sigma_t)\\ \nonumber
 && +\gamma_a \lambda_a \sin(\frac{\pi}{3}\sigma_t+\theta_z \sigma_h) - \gamma_b \lambda_b \sin(-\frac{\pi}{3}\sigma_t+ \theta_z \sigma_h) \\ \nonumber 
& = & \frac{\sqrt{3}}{2}[\gamma_{1a}\lambda_a+\gamma_{1b}\lambda_b+\gamma_a \lambda_a \cos(\theta_z)+\gamma_b\lambda_b \cos(\theta_z)]\sigma_t \\ \nonumber
&& + \frac{1}{2}\sin(\theta_z)(\gamma_b\lambda_b-\gamma_a\lambda_a)\sigma_h 
\end{eqnarray}
where $\theta_z=2\pi |q_z|$, $\lambda_a=\cos(\frac{\pi}{6}+\theta_0)$,
$\lambda_b=\cos(\frac{\pi}{6}-\theta_0)$, and $\theta_0$ is the ferroaxial
rotation angle of tetrahedron. Since $\gamma_b-\gamma_a$ corresponds to $A$, here we successfully reproduce the phenomenological
relation $P_c=c_1\sigma_t+c_2A\sigma_h$.
Based on the magnititude of the corresponding $J'$s, we expect that
$\gamma_b-\gamma_a$ should be small in RFMO, and so will be
the contribution to the polarization from the $\sigma_h$ term. In addition, the ferroaxial term is proportional to 
$\sin(\theta_z)=\sin(0.44 \times 2\pi)=0.37$ in experiment, much smaller than $|\cos(\theta_z)|=0.93$ in the $\sigma_t$ term, which further reduces the relative contribution from the ferroaxial term. We predict that the ferroaxial term will be larger in system with a shorter $c$ axis and a greater ferroaxial rotation, which can both be achieved by a combination of uniaxial and chemical pressure. The former enhances the diagonal coupling, while the latter increases both the difference between $\gamma_a$ and $\gamma_b$ and the spin rotation angle $\theta_z$.  This scenario would produce much larger values of $P$, as realised in CaMn$_7$O$_{12}$,\cite{johnson12,natasha12} where the spin rotation angle $\theta_z$ is much greater (124$^{\circ}$) and much stronger nearest-neighbour exchange provides the diagonal interactions. 

\section{Conclusions}
Through first-principles calculations and analysis of a Heisenberg model, we demonstrate that the ferroaxial structure
in RFMO does indeed induce the incommensurate helical spin structure
along the $c$ direction via the symmetric exchange interactions. 
We find the electric polarization resulting from the ferroaxial term to
be negligible (i.e., $c_2 \approx$0) compared to the contribution from the triangular chirality.
The electric polarization in RFMO arises as a secondary effect of a structural distortion induced through the inverse DM mechanism.
We further give an explicit microscopic expression for polarization based on this mechanism and explain why the ferroaxial contribution is small in RFMO.

\acknowledgments
The work done at the University of Oxford was funded by an EPSRC grant, number EP/J003557/1, entitled ``New Concepts in Multiferroics and Magnetoelectrics".  LH acknowledges the support from the Chinese National Fundamental Research Program 2011CB921200, the National Natural Science Funds for Distinguished Young Scholars, and CNSF Grant No. 11374275.

{\it Note added}.\textemdash After we submitted the paper, there is an experiment Ref.~\onlinecite{white13} that has measured the effective spin exchange interactions, and obtained $J_1$=0.086 meV, and $J_p$= 0.0007 meV. In our calculations, we normalized S to 1, while the experimental paper use S=5/2. When rescale our calculated J's from U=4 eV using S=5/2, we have $J_1$ = 0.12 meV, $J_2$= 0.0028 meV, which are in overall good agreement with the experimental values.


\end{document}